\documentclass[copyright,creativecommons]{eptcs}
\providecommand{\event}{FMSPLE 2015}

\usepackage{color}
\usepackage{listings}
\newcommand{\lline}[1]{Line~\ref{#1}} % ref line of listings
\newcommand{\llines}[2]{Lines~\ref{#1}-\ref{#2}} % ref lines of listings
\usepackage{multirow}
\usepackage{colortbl}
\usepackage{booktabs}

\usepackage{amssymb,amsmath,amstext,latexsym,marvosym,xfrac}
\usepackage{graphicx}
\usepackage{times}
\newcommand{\irule}[2]{\protect{\frac{\displaystyle\rule[-1ex]{0cm}{3ex}#1}{\displaystyle\rule[-.5ex]{0cm}{3ex}#2}}}
\newcommand{\arco}[2]{^{\ \underrightarrow{_{\ \ {#2} \ \ }}} \ }
\newcommand{\flan}{\textsc{FLan}}
\newcommand{\pflan}{\textsc{PFLan}}
\newcommand{\nat}{\mathbb N}
\newcommand{\real}{\mathbb R}
\newcommand{\has}[1]{\mathit{has}(#1)}
\newcommand{\done}[1]{\mathit{do}(#1)}

\newcommand{\req}[2]{#1 \triangleright #2}
\newcommand{\xor}[2]{#1 \otimes #2}
\newcommand{\install}[1]{\mathsf{install}(#1)}
\newcommand{\ask}[1]{\mathsf{ask}(#1)}
\usepackage{cite,url,breakurl}
\usepackage[T1]{fontenc}
\providecommand{\sortnoop}[1]{}
\usepackage{subfigure}
\title{Quantitative Analysis of Probabilistic Models of Software Product Lines with Statistical Model Checking
}
\author{%
  Maurice H. ter Beek
  \institute{ISTI--CNR, Pisa, Italy}
  \and
  Axel Legay
  \institute{Inria, Rennes, France}
  \and
  Alberto Lluch Lafuente
  \institute{DTU, Lyngby, Denmark}
  \and
  Andrea Vandin
  \institute{U Southampton, UK}
}

\def\titlerunning{Quantitative Analysis of Probabilistic SPL Models with Statistical Model Checking}
\def\authorrunning{M.H. ter Beek, A. Legay, A. Lluch Lafuente \& A. Vandin}

\begin{document}

\maketitle

\begin{abstract}
We investigate the suitability of statistical model checking techniques for analysing quantitative properties of software product line models with probabilistic aspects.  
For this purpose, we enrich the feature-oriented language \flan{} with action rates, which specify the likelihood of exhibiting particular behaviour or of installing features at a specific moment or in a specific order. The enriched language (called \pflan{}) allows us to specify models of software product lines with probabilistic configurations and behaviour, e.g.\ by considering a \pflan{} semantics based on discrete-time Markov chains.  
The Maude implementation of \pflan{}
% which extends the one of \flan{}, 
is combined with the distributed statistical model checker MultiVeStA to perform quantitative analyses of a simple product line case study. The presented analyses include the likelihood of certain behaviour of interest (e.g.\ product malfunctioning) and the expected average cost of products. 
\end{abstract}

% !TEX root = FMSPLE.tex

\section{Introduction}
\label{sec:intro}

The modelling and analysis by means of process calculi and formal verification techniques like model checking of the variety of configurations and behaviour that is common to a software product line (SPL) is gaining momentum~\cite{EW11,BMS12,BLP13,CCHLS12,CCSHLR13,CCHLS14,BDV14a,BDV14b,SB14,TAKSS14,Tri14,BFGM15}. Compared to the complexity of verifying the behaviour of a single product or a single system, the variability inherent to SPL adds another dimension as the number of possible products of an SPL may be exponential in the number of features~\cite{BCLW13}.
In~\cite{BLP13}, we introduced the feature-oriented language \flan{} as a high-level modelling language for SPLs. A rich set of process-algebraic operators allows one to specify in a procedural, operational way both the configuration and the behaviour of products, while a constraint store allows one to specify in a declarative way all common structural constraints known from feature models and additional action constraints typical of feature-oriented software development. On the one hand, the execution of a process is constrained by the store (e.g.\ to avoid introducing inconsistencies), while on the other hand a process can query the store (e.g.\ to resolve configuration options) or update the store (e.g.\ to add new features, also at run time or by means of a staged configuration process). An implementation of \flan{} in the executable modelling language Maude~\cite{CDELMMT07} allows one to exploit Maude's rich toolkit for a variety of formal analyses of \flan{} models, ranging from consistency checking (by means of SAT solving) to model checking. 

In this paper, we introduce a probabilistic extension of \flan{}: \pflan{} allows to equip actions with rates to specify probabilistic SPL models (e.g.\ to model uncertainty, failure rates, randomisation). This paves the way for quantitative analyses (e.g.\ to measure quality of service, reliability, performance). Here we present a proof-of-concept use of an implementation of \pflan{} in Maude in combination with the distributed statistical model checker MultiVeStA~\cite{SV13} to estimate the likelihood of specific behaviour. Formally, our approach is to perform a sufficient number of probabilistic simulations of a \pflan{} model to obtain statistical evidence (with a desired level of statistical confidence) of quantitative properties under scrutiny. The properties are formulated in MultiVeStA's property specification language MultiQuaTEx, which allows to express and evaluate more than one property over the same simulated path (behaviour)~\cite{SV13}. 
The advantage over exhaustive (probabilistic) model checking is that there is no need to generate entire state spaces. We argue that this outweighs the main disadvantage of having to give up on obtaining exact results (100\% confidence) with exact analysis techniques like probabilistic model checking, in particular when examining an SPL, given their possibly exponential number of products.

We refer to~\cite{BK08} for (probabilistic) model checking and to~\cite{LDB10,LL14} for statistical model checking. An overview of related work on applying formal analysis techniques in SPLE can be found in~\cite{TAKSS14}, while~\cite{BLP13} contains an extensive discussion of related work on model-checking SPL behaviour. 
As far as we know, there are only a few, quite different, approaches on probabilistic model checking of an SPL~\cite{GS13,VK13,DKB14}, whereas we present here the first application of statistical model checking in SPL engineering (SPLE).

%\noindent
The paper outline is as follows. 
Section~\ref{sec:example} contains a toy example of a product line of coffee machines, adapted from~\cite{BFGM15,BLP13,BMS12,BDV14a,BDV14b}. 
Section~\ref{sec:language} presents \pflan{}, followed by a \pflan{} model of the example in Section~\ref{sec:model}.
MultiVeStA is introduced in Section~\ref{sec:multivesta}, followed by experimental quantitative analyses of the example in Section~\ref{sec:analysis}. 
Section~\ref{sec:discussion} summarises the contributions of this paper and discusses future work.

\setlength{\belowcaptionskip}{20pt}
% !TEX root = FMSPLE.tex

\section{An Example Product Line of Coffee Machines}
\label{sec:example}

Our toy example is a (simplistic) product line of coffee machines with %which is quite popular in SPLE~\cite{FG08,ABFG10b,CHSLR10,ABFG11b,MPC11,BLP13,CCSHLR13,CCHLS14,BDV14b}. We consider 
the following list of requirements:
\begin{enumerate}%\itemsep=-1pt\parsep=-1pt
\item Initially, a coin must be inserted: either a euro, exclusively for  products for the European market, or a dollar, exclusively for  Canadian products; 
\item An optional cancel button allows the user to cancel coin insertion, after which the coin is returned;  %and returns idle;
\item A machine that contains a coin must offer a choice to add sugar, followed by a choice of beverages; 
\item The choice of drinks (coffee, tea, cappuccino) varies, but all products must offer at least one drink, tea may be offered only by European products, and products offering cappuccino must offer coffee; 
\item An optional ringtone may be rung after beverage delivery. It must be rung after serving cappuccino; 
\item After the drink is taken, the machine returns idle.
\end{enumerate}
These requirements for products combine structural constraints defining valid feature configurations (e.g.\ \emph{\lq\lq every product must offer at least one beverage\rq\rq}) with temporal constraints defining valid product behaviour in terms of valid action sequences (e.g.\ \emph{\lq\lq a ringtone must be rung after serving a cappuccino\rq\rq}). 

%\begin{figure}
%\centering\includegraphics[width=\textwidth]{FeatureModelX}
%\caption{\label{fig:FM}Feature model of Coffee Machine (with shorthand names)}
%\end{figure}

\begin{figure}[t!]
\centering\includegraphics[width=.9\textwidth]{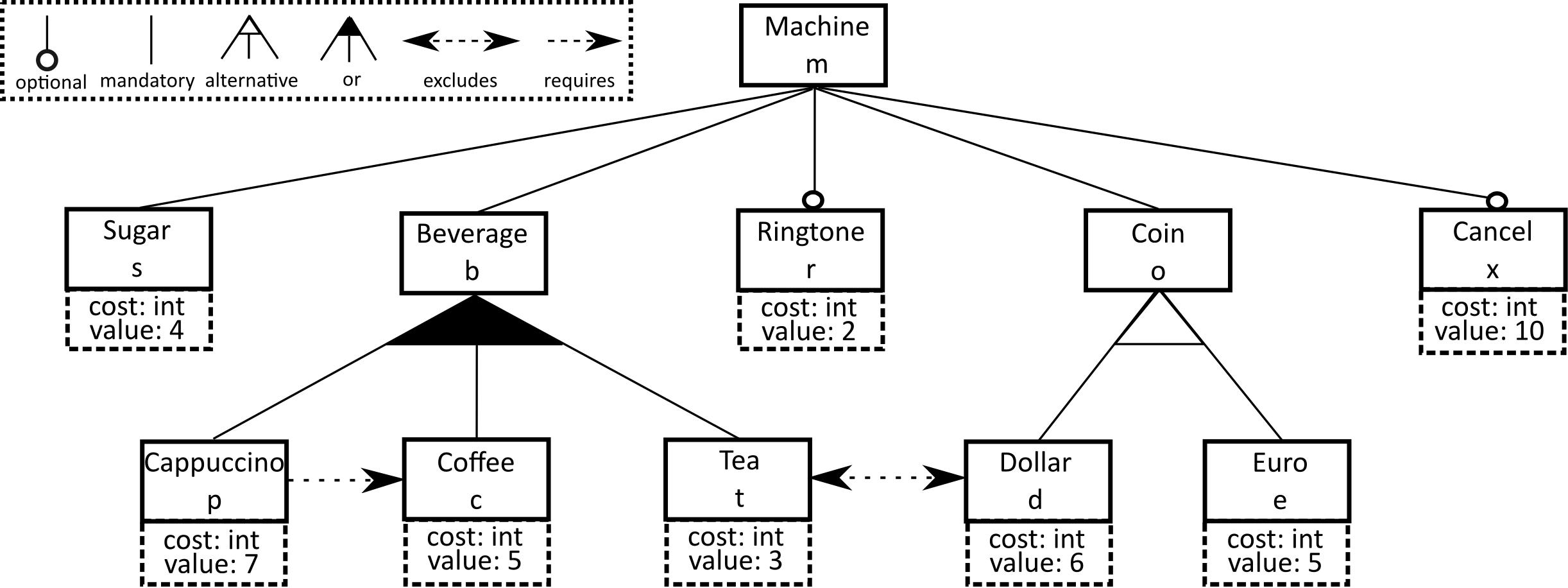}
\caption{\label{fig:AFM}Attributed feature model of Coffee Machine (with shorthand names)}
\end{figure}

The de facto standard variability model in SPLE is a \emph{feature model\/}~\cite{KCHNP90,SHT06}. It provides a compact representation of all valid products of a product line in terms of their features (behaviour is not captured). An (attributed) feature model of our example is depicted in Fig.~\ref{fig:AFM}. 
%Graphically, features are nodes of a tree, with the family as its root, and the following relations between features represent constraints (the first four structure the tree, the latter two model cross-tree constraints):
%%
%\begin{description}
%\item[Optional] features: may be present only if their parent is;
%\item[Mandatory] features: must be present iff their parent is;
%\item[Or] features: at least one must be present if their parent is;
%\item[Alternative (xor)] features: only one must be present if their parent is;
%\item[Requires] relation among two features: one's presence implies that of the other;
%\item[Excludes] relation among two features: their presence is  mutually exclusive.
%\end{description}
%%
%The feature model in Fig.~\ref{fig:FM}
It has a root (feature) $m$ and a set of non-trivial features, partitioned into the sets $\{\mathit{b},\mathit{o}\}$ of \emph{compound features\/} and $\mathit{Features} = \{\mathit{s},\mathit{r},\mathit{x},\mathit{p},\mathit{c},\mathit{t},\mathit{d},\mathit{e}\}$ of \emph{primitive features\/}.\footnote{In case no confusion can arise, we often simply speak of features when we actually refer to the primitive features.} The only purpose of the former is to group the (primitive) features in the tree, whereas the latter define user observable configuration parameters~\cite{Bat05,SHT06}.
We identify a product from the product line with a non-empty subset of \textit{Features}. Deciding whether a product satisfies a feature model can be reduced to Boolean satisfiability (SAT), and efficiently be computed with SAT solvers~\cite{Bat05}.

By equipping features with (non-functional) attributes (e.g.\ $\mathit{cost}(\mathit{Tea}) = 3$) we obtain an \emph{attributed feature model\/}.%An attributed feature model of our example is depicted in Fig.~\ref{fig:AFM}. 
\footnote{
Additional quantitative constraints on (combinations of) %compound or primitive 
features may be defined
(e.g.\ %$\mathit{cost}(\mathit{Beverage})\leq 10$, 
$\mathit{cost}(\mathit{Sugar}) + \mathit{cost}(\mathit{Ringtone})\leq\mathit{cost}(\mathit{Coin})$) %or $\mathit{cost}(\mathit{Machine})\leq 35$) 
but we prefer to neglect them in this paper, as such constraints require the use of SMT solvers like Microsoft's Z3~\cite{MB08}, currently under integration in our framework.
%
%This naturally leads to the use of the computationally more expensive Satisfiability Modulo Theory (SMT) solvers like Microsoft's Z3~\cite{MB08}. In the future, we intend to integrate this into our approach by implementing Microsoft's efficient Z3 SMT/SAT constraint solver~\cite{MB08} in Maude.}
% scould also be defined.
}
The cost function $\mathit{cost}: \mathit{Features}\to\nat$, associated to the attribute $\mathit{cost}$, straightforwardly extends to %compound features and 
products: $\mathit{cost}(\mathit{product}) = \textstyle{\sum}\,\{\,\mathit{cost}(\mathit{feature})\mid \mathit{feature}\in\mathit{product}\,\}$. %(and likewise for compound features).  
Thus, intuitively, $\mathit{cost}$ can be seen as a labelling function assigning a non-negative integer to each product defined by a feature model. %, after which the valid products are those that satisfy the quantitative constraints.
%
%Ignoring attributes and cross-tree constraints, the feature model in Fig.~\ref{fig:AFM} yields $(2^{3}-1)\times 2\times 2\times 2 = 56$~valid products (since $\mathit{Sugar}$ is mandatory and a product without any beverage is invalid) out of the $2^{8} - 1$ possible non-empty sets of concrete features. The cross-tree constraints reduce this number to $28$.
%, while it is further reduced to $27$~valid products if the attributes plus the additional quantitative constraint $\mathit{cost}(\mathit{Machine})\leq 35$ are considered (since $\textit{cost}(\{s,r,x,p,c,t,e\}) = 36$).

%Feature attributes typically are not Boolean~\cite{CSHL13}. 
%Therefore, the problem of deciding whether or not a product satisfies an attributed feature model requires more general satisfiability-checking techniques than mere SAT solving. 
%This naturally leads to the use of the computationally more expensive Satisfiability Modulo Theory (SMT) solvers like Microsoft's Z3~\cite{MB08}. In the future, we intend to integrate this into our approach by implementing Microsoft's efficient Z3 SMT/SAT constraint solver~\cite{MB08} in Maude.}

\setlength{\belowcaptionskip}{10pt}
% !TEX root = FMSPLE.tex

\section{PFL\textbf{\textsc{an}}: Syntax and Semantics}
\label{sec:language}

The feature-oriented language \pflan{} is a probabilistic extension of \flan{}~\cite{BLP13}, a process algebra that neatly separates declarative (pre-)configuration from procedural run-time aspects. \pflan{} is inspired by the concurrent constraint programming paradigm of~\cite{SR90}, its adoption in process calculi~\cite{BM07} and its stochastic extension~\cite{Bor06}. A constraint store allows one to specify all common constraints known from feature models 
%, as well as additional quantitative constraints (such as the cost of a feature or of a configuration)  
in a \emph{declarative\/} way, while a rich set of process-algebraic operators allow to specify the configuration and behaviour of product lines in a \emph{procedural\/} way. The semantics smoothly unifies \emph{static\/} (pre-configuration) and \emph{dynamic\/} (run-time) feature selection.

The core notions of \pflan{} are \emph{features\/}, \emph{constraints\/}, \emph{processes\/} (with action \emph{rates\/}) and \emph{fragments\/}, all visible in its syntax in Fig.~\ref{fig:syntax}. 
%\marginpar{\mbox{add costs /} \mbox{quantitative} \mbox{constraints} to syntax?}
More precisely, $f$ and~$g$ range over features while the syntactic categories~$F$, $S$ and~$P$ correspond to fragments, a constraint store and processes (with actions from $A$), respectively. 

%\paragraph*{Features} 

%A \emph{feature\/} is a term describing specific elements or properties of a product. 
The universe of (primitive) features is denoted by~$\mathcal{F}$. The features of our example are the accepted coin slots (i.e.\ \textit{euro\/} and \textit{dollar}), the offered drinks (i.e.\ \textit{coffee}, \textit{tea\/} and \textit{cappuccino}) and the additional capabilities \textit{sugar\/}, \textit{cancel\/} and \textit{ringtone\/} (to add sugar, cancel coin insertion and ring a tone, respectively). %Our approach is general enough to accommodate all common notions of features~\cite{CHS08}.

%\paragraph*{Constraints} 

\medskip
The declarative part of \pflan{} is represented by a store of
%constraint store which contains %(quantitative) 
constraints on features extracted from the product line requirements plus some additional information (e.g.\ about the context wherein the product will operate).  
Two important notions of a constraint store $S$ are the \emph{consistency\/} of~$S$, denoted by $\emph{consistent(S)}$ (which in our case amounts to logical satisfiability of all constraints constituting~$S$) and the \emph{entailment\/} $S\vdash c$ of constraint~$c$ in~$S$ (which in our case amounts to logical entailment).  
A constraint store contains any term generated by~$S$ according to the syntax of \pflan{}. %grammar of \pflan{} in Fig.~\ref{fig:syntax}. 
The most basic constraint stores are $\top$~(no constraints at all), $\bot$~(inconsistent constraints) and ordinary Boolean propositions (generated by~$K$).  
Constraints can be combined by juxtaposition (its semantics amounts to logical conjunction). 

We assume that constraints on features are expressed using Boolean propositions (cf.~\cite{SHT06}). 
Moreover, we assume that the universe~$\mathcal{P}$ of propositions contains a Boolean predicate $\has{f}$ that can be used to denote the presence of a feature $f$ in a product.  
Boolean propositions can also be used to represent additional information such as contextual facts. 
In our example we use the Boolean propositions $\mathit{in(Europe)}$ and $\mathit{in(Canada)}$ to state the fact that the coffee machine being configured is meant to be used in Europe or in Canada, respectively.  
Finally, Boolean propositions can state relations between contextual information and features, like $\has{\textit{euro}}\rightarrow in(Europe)$ (i.e.\ a coffee machine has a coin slot for euro's only if it is intended for the European market).

\begin{figure}
\hrulefill 
\vspace*{-0.25cm}
$$
\begin{array}{rcl}
\mathit{F} & ::= & [S\mid P] \\[0.25em]
% Features Constraints
S, T & ::= & K \, \mid \, \req{f}{g} \, \mid \, \xor{f}{g} \, \mid  \, S \ T \, \mid \, \top \, \mid \, \bot \\[0.25em]
% Processes
P, Q & ::= & \emptyset \, \mid \, \mathit{X} \, \mid \, (A,r).P \, \mid \, P + Q \, \mid \, P ; Q \, \mid \, P\parallel Q \\[0.25em]
%\end{array}
%\hspace{1cm}
%\begin{array}{rcl}
% Actions
A & ::= & a \, \mid \, \install{f} \, \mid \, \ask{K} \\[0.25em]
% Contexts and Context Conditions
\mathit{K} & ::= & p \, \mid \, \neg\,K \, \mid \, K\,\vee\,K 
%\phantom{A}
\end{array}
$$

\vspace*{-0.25cm}
\hrulefill
\caption{\label{fig:syntax}Syntax of \pflan{}, where $r\in\real^+$, $a\in\mathcal{A}$, $p\in\mathcal{P}$ and $f,g\in\mathcal{F}$}
\end{figure}

\noindent
Two common cross-tree constraints are instead handled as first-class citizens in \pflan{}. 
A constraint $\req{f}{g}$ expresses that feature~$f$ requires the presence of feature~$g$, whereas a constraint $\xor{f}{g}$ expresses that features~$f$ and~$g$ mutually exclude each other's presence (i.e.\ they are incompatible).  
Also these constraints could of course be encoded as Boolean propositions (e.g.\ $\xor{f}{g}$ and $\req{f}{g}$ can equivalently be expressed as $\has{f} \leftrightarrow \neg \has{g}$ and $\has{f} \rightarrow \has{g}$, respectively).
We in fact use such logical encodings to reduce consistency checking and entailment to logical satisfiability (and hence exploit Maude's SAT solver). 
However, we prefer to keep this first-class treatment as syntactic sugar. 
In our example, we extract $\xor{\mathit{dollar}}{\mathit{euro}}$ to formalise that $\mathit{euro}$ and $\mathit{dollar}$ are mutually exclusive features (requirement~1) and $\req{\mathit{cappuccino}}{\mathit{coffee}}$ to formalise that $\mathit{cappuccino}$ requires $\mathit{coffee}$ (requirement~3).

We also consider a class of \emph{action constraints\/}, reminiscent of featured transition systems (FTS)~\cite{CCSHLR13}. In an FTS, transitions are labelled with actions and with Boolean constraints over the set of features. We associate arbitrary constraints to actions rather than to transitions (and we moreover add a rate to the actions, discussed below). 
In a coffee machine offering coffee, e.g., we will use $\mathit{coffee}$ for the (user) action of choosing coffee and $\done{\mathit{coffee}}$ as a proposition stating the execution of that action. The relation between the action $\mathit{coffee}$ and the presence of the corresponding feature $\mathit{coffee}$ can be formalised as $\done{\mathit{coffee}}\rightarrow\has{\mathit{coffee}}$, i.e.\ the choice for coffee requires coffee being offered by the coffee machine.  
In general, we assume that each action~$a$ may have a constraint $\done{a}\rightarrow p$, where $p\in\mathcal{P}$ is a proposition. Such constraints act as a kind of guards to allow or forbid the execution of actions (cf.\ the discussion of the rule \textsc{Act} below). Note that these action constraints could also be more complex, e.g.\ we could define an action $\mathit{caf\!\acute{e}}$-$\mathit{au}$-$\mathit{lait}$ together with the action constraint $\done{\mbox{$\mathit{caf\!\acute{e}}$-$\mathit{au}$-$\mathit{lait}$}}\rightarrow(\has{\mathit{coffee}}\wedge\has{\mathit{milk}})$.

%\paragraph*{Processes} 

%\marginpar{~ \ ~\\ add costs!}
\medskip
The procedural part of \pflan{} is represented by \emph{processes\/} which can be of the following type: 
\begin{description}%\itemsep=-1pt\parsep=-1pt
\item [\quad\ $\emptyset$]
%$\emptyset$, i.e. 
the empty process that does nothing; 
\item [\quad\ $\mathit{X}$] 
%$\mathit{X}$, i.e. 
a process identifier;\footnote{We assume there is a set of process definitions of the form $\mathit{X}\doteq P$ and recursively defined processes to be finitely branching.} %. We moreover assume that recursively defined processes are finitely branching, which can be ensured in standard ways (e.g. prefixing every occurrence of a process identifier~$\mathit{X}$ with an action or constraining process definitions to be of the form $\mathit{X} \doteq A.P$).} 
\item [\quad\ $(A,r).P$]
%$(A,r).P$, i.e. 
a process that can perform action~$A$ with rate~$r$ and then behaves as~$P$; 
\item [\quad\ $P + Q$] 
%$P + Q$, i.e. 
a process that can non-deterministically choose to behave as either~$P$ or~$Q$; 
\item [\quad\ $P ; Q$] 
%$P ; Q$, i.e. 
a process that must progress first as~$P$ and then as~$Q$; 
\item [\quad\ $P\parallel Q$]
%$P\parallel Q$, i.e. 
a process formed by the parallel composition of~$P$ and~$Q$, which evolve independently.  
\end{description}
% Actions and features
We distinguish ordinary actions from a universe~$\mathcal{A}$ and two special actions $\install{f}$ (which will be used to denote the dynamic installation of a feature $f$) and $\ask{K}$ (which can used to query the store for the validity of a Boolean proposition from $K$). 
As we will see shortly, each action type is treated differently in the operational semantics. 
%Such first-class treatment is useful, among other things, to present a more clear semantics of feature installation. 
Note, moreover, that each action has an associated \emph{rate\/} (sometimes called \emph{weight\/}),
which is used to determine the probability that this action is executed. As usual, the probability to execute an action in a certain state depends on the rates of all other actions enabled in the same state. We will illustrate this in our example in Section~\ref{sec:model}. 
We consider the actions $\mathit{euro}$, $\mathit{dollar}$ (respective coin insertion), $\mathit{cancel}$ (cancellation of coin insertion), $\mathit{sugar}$ (sugar selection), $\mathit{ringtone}$ (ringtone emission), $\mathit{coffee}$, $\mathit{tea}$ and $\mathit{cappuccino}$ (beverage selection) in our example. Their associated rates are discussed below. For simplicity we consider only constant rates, but our framework can be easily extended to allow store-dependent rates (e.g.\ to be able to reflect a higher probability to order cappuccino in Europe). 

%\paragraph*{Fragments} 
%
Finally, a \emph{fragment\/}~$F$ is a term $[S\mid P]$, composed by a constraint store~$S$ and a process~$P$. 
%\marginpar{\mbox{note how} \mbox{store may} influence behaviour}
These two components may influence each other according to the concurrent constraint programming paradigm~\cite{SR90}: a process may update its store which, in turn, may condition the execution of the process' actions. 

%\paragraph*{Semantics of Fragments} 

\begin{figure}[t]
\hrulefill 
\vspace*{-0.25cm}
$$
\begin{array}{ccc}
(\textsc{Inst})\, \irule{\mathit{consistent}(S\ \has{f})}{[S\mid(\install{f},r).P]\arco{\install{f}}{r} [S\ \has{f}\mid P]} &\quad& (\textsc{Or})\, \irule{[S\mid P]\arco{\ell}{r} [S'\mid P']}{[S\mid P + Q]\arco{\ell}{r} [S'\mid P']}\\[1.5em]
(\textsc{Ask})\, \irule{S\vdash K}{[S\mid(\ask{K},r).P]\arco{\ask{K}}{r} [S\mid P]} &\quad& (\textsc{Seq})\, \irule{[S\mid P]\arco{\ell}{r} [S'\mid P']}{[S\mid P; Q]\arco{\ell}{r} [S'\mid P'; Q]}\\[1.5em]
(\textsc{Act})\, \irule{S = (\done{a}\rightarrow K)\qquad S\vdash K}{[S\mid(a,r).P]\arco{a}{r} [S\mid P]} &\quad& (\textsc{Par})\, \irule{[S\mid P]\arco{\ell}{r} [S'\mid P']}{[S\mid P\parallel Q]\arco{\ell}{r} [S'\mid P'\parallel Q]}
\end{array}
$$

\vspace*{-0.25cm}
\hrulefill
\caption{\label{fig:semantics}Reduction semantics of \pflan{}%, where $S\vdash K$ denotes $K$ is derivable from $S$
} 
\end{figure}

\begin{figure}
\hrulefill 
\vspace*{-0.25cm}
$$
\begin{array}{rclcrclcrcl}
P + (Q + R) &\equiv& (P + Q) + R &\quad& P + \emptyset &\equiv& P &\quad& P + Q &\equiv& Q + P\\[0.15em]
P\parallel (Q\parallel R) &\equiv& (P\parallel Q)\parallel R &\quad&  P\parallel \emptyset &\equiv& P &\quad& P\parallel Q &\equiv& Q\parallel P\\[0.15em]
P ; (Q ; R) &\equiv& (P ; Q) ; R &\quad& P ; \emptyset &\equiv& P \ \ \, \equiv \, \ \ \emptyset ; P &\quad& P &\equiv & P[^Q/_X] \text{ \ if \ } \mathit{X} \doteq Q
\end{array}
$$

\vspace*{-0.25cm}
\hrulefill
%\caption{Reduction semantics (top) and structural congruence (bottom) of \pflan{}}%, where $S\vdash K$ denotes $K$ is 
\caption{\label{fig:congruence}Structural congruence in \pflan{}}
%\label{fig:semantics}\label{fig:congruence}
\end{figure}

\medskip
The operational semantics %of closed fragments (i.e.\ its reduction semantics) 
is formalised in terms of the state transition relation $\rightarrow\,\subseteq \nat^{\mathbb{F}\times\real^+\times \mathbb{F}}$ defined in Fig.~\ref{fig:semantics}, where~$\mathbb{F}$ denotes the set of all terms generated by~$F$ in the grammar of Fig.~\ref{fig:syntax}. Note that multisets of transitions are needed to deal with the possibility of having multiple instances of a transition $F \xrightarrow{r} G$. 
%\marginpar{update \mbox{semantics}}
Technically, such a reduction relation is defined in structural operational semantics (SOS) style (i.e.\ by induction on the structure of the terms denoting a fragment) modulo the structural congruence relation $\equiv\,\subseteq\mathbb{F}\times\mathbb{F}$ defined in Fig.~\ref{fig:congruence}. The reduction relation implicitly defines a labeled transition system LTS, whose labels are rates.  
%
% Questo discourse di sotto può creare confusione. Meglio levarlo. 
%At this point, it is worth to note that we might actually obtain a multi-labelled transition system, since the semantics might generate more than one transition with same label among two states. Consider, e.g., a store~$S$ that allows the execution of the actions $a_1$ and $a_2$ (which do not modify the store) and a process $P= (a_1,r).\emptyset + (a_2,r).\emptyset$. Applying the rules of Fig.~\ref{fig:semantics} to $[S\mid P]$ results in a LTS with  states $[S\mid P]$ and $[S\mid \emptyset]$ and two distinct transitions labelled with rate $r$ from the former state to the latter one. 
%
It is straightforward to obtain a discrete time Markov chain (DTMC) from such LTSs by normalising the rates into $[0..1]$ such that in each state, the sum of the rates of its outgoing transitions equals one. %In the example above, we would obtain rate $\frac{r}{r+r} = \sfrac{1}{2}$ for both transitions.
As usual, in the resulting DTMC the label of a transition  corresponds to the probability that such a transition is executed starting from its source state.
Recall that we advocate the use of statistical model checking because in general the DTMC is too large to generate.
%
%The reduction rules in Fig.~\ref{fig:semantics} are expressed in terms of a (possibly empty) set of premises (above the line) and a conclusion (below the line).

The rules \textsc{Inst} and \textsc{Act} of the semantics are very similar, both allowing a process to execute an action if certain constraints are satisfied. 
Rule \textsc{Inst} forbids inconsistencies caused by the introduction of new features. 
It can be seen as a particular instance of the rule for the $\textsf{tell}$ operation of concurrent constraint programming~\cite{SR90} instantiated as $\textsf{tell}(\emph{has}(f))$. 
%
%\marginpar{rewrite and update}
%
Rule \textsc{Act} forbids inconsistencies with respect to action constraints. 
A typical action constraint is $\done{a}\rightarrow \has{f}$, i.e.\ action $a$ is subject to the presence of feature $f$. 
Rule \textsc{Ask} formalises the semantics of the $\ask{\cdot}$ operation from concurrent constraint programming~\cite{SR90}. 
It allows a process to be blocked until a proposition can be derived from the store. 
Rules \textsc{Par}, \textsc{Seq} and \textsc{Or} formalise interleaving parallel composition, sequential composition and non-deterministic choice, respectively. 
%
%This non-determinism can be solved procedurally (by relying on $\ask{\cdot}$ actions) or declaratively (by having it solved by the constraint store), thus providing designers a lot of flexibility (as illustrated in~\cite{BLP13}). 
%
%Also rules \textsc{Seq} and \textsc{Par} are standard. The former formalises the usual sequential composition, while the latter formalises an interleaving parallel composition. 
Note that the non-determinism introduced by choices and parallel composition is probabilistically resolved in the aforementioned DTMC semantics. 

\enlargethispage*{\baselineskip}
Summarising, we note a variety of ways in which a feature~$f$ can be included in a configuration. 
First, an \emph{explicit\/} and \emph{declarative\/} way is to include the proposition $\has{f}$ in the initial store; this is the way to include core features.
Second, an \emph{implicit\/} and \emph{declarative\/} way is to derive~$f$ from other constraints  
%This is the way to include features that are not known as core features, but that turn out to be enforced by the constraints 
(e.g.\ if a store contains $\req{g}{f} $ and $\has{g}$, then $f$'s presence follows). 
Third, a \emph{procedural\/} way is to dynamically install~$f$ at run time; this key aspect originating from \flan{} enables staged configuration as known from dynamic software product lines~\cite{CHE04,BLLBGS14}.
Building on \flan{}, \pflan{} combines these three ways in an elegant and consistent manner. 
The introduction of action rates in \pflan{} moreover allows one to specify  probabilistic aspects of SPL models such as the behaviour of the user of a product and the likelihood of installing a certain feature at a specific moment with respect to that of other features.

\setlength{\belowcaptionskip}{20pt}
% !TEX root = FMSPLE.tex

%\paragraph*{Example}
\section{A PFL\textbf{\textsc{an}} Model of the Example Product Line}
\label{sec:model}
%\paragraph*{A \pflan{} Model of the Example Product Line.}
%
Fig.~\ref{fig:coffee} shows a specification of the family of coffee machines in \pflan{}.
Fragment~$\mathit{F}$ is composed of store~$\mathit{S}$ and a process~$\mathit{Q}$. The latter specifies an initial configuration phase~$\mathit{D}$, during which all primitive features except $\mathit{ringtone}$ can be installed (the order of installation is influenced by the relative weight of the features, more on this below). This phase ends at a certain moment when a specific product (coffee machine) is said to be pre-configured, modeled by the installation of an ad-hoc defined feature $\emph{pre-conf}$, thus initiating the execution of process~$\mathit{R}$, which specifies the product's run-time behaviour. Note that it is specifically allowed to install (or bind) a feature at run time (cf.\ $\mathit{ringtone}$ in our toy example). 

\begin{figure}[tb!]
\small\centering
\hrulefill 
$$
\begin{array}{rcl}
\textit{F} & \doteq & [\textit{S}\mid \textit{Q}]\\[0.25em]
\textit{S} & \doteq & \textit{S}_1 \ \textit{S}_2 \\[0.2em] %\ \textit{S}_3\\
\textit{S}_1 & \doteq & 
%\text{/* Basic structural constraints */}\\ 
% at least one of euro or dollar
\has{\mathit{euro}}\vee\has{\mathit{dollar}} \quad 
% euro, exclusively for products for the European market
\has{\mathit{euro}}\rightarrow\mathit{in(Europe)} \quad 
% dollar, exclusively for Canadian products
\has{\mathit{dollar}}\rightarrow\mathit{in(Canada)} \\[0.15em]
% every product must offer at least one beverage
&& \has{\mathit{coffee}}\vee\has{\mathit{tea}}\vee\has{\mathit{cappuccino}} \qquad
% tea may be offered only by European product
\has{\mathit{tea}}\rightarrow\mathit{in(Europe)} \\[0.15em]
% either a euro, or a dollar
%&& \text{/* Cross-tree constraints */} \\ 
&& \xor{\mathit{dollar}}{\mathit{euro}} \qquad
% all products that offer cappuccino must also offer coffee
\req{\mathit{cappuccino}}{\mathit{coffee}} \\[0.15em]
% action constraints
&& \done{\mathit{euro}}\rightarrow\has{\mathit{euro}} \qquad
\done{\mathit{dollar}}\rightarrow\has{\mathit{dollar}} \\[0.15em]
&& \done{\mathit{sugar}}\rightarrow\has{\mathit{sugar}} \qquad 
\done{\mathit{ringtone}}\rightarrow\has{\mathit{ringtone}} \qquad \done{\mathit{cancel}}\rightarrow\has{\mathit{cancel}} \\[0.15em] 
&& \done{\mathit{pour\_sugar}}\rightarrow\has{\mathit{sugar}} \qquad \done{\mathit{out\_of\_sugar}}\rightarrow\has{\mathit{sugar}} \\[0.15em]
&& \done{\mathit{return\_coin}}\rightarrow\has{\mathit{cancel}} \qquad \done{\mathit{no\_return}}\rightarrow\has{\mathit{cancel}} \\[0.15em]
&& \done{\mathit{coffee}}\rightarrow\has{\mathit{coffee}} \qquad
\done{\mathit{tea}}\rightarrow\has{\mathit{tea}} \qquad\,
\done{\mathit{cappuccino}}\rightarrow\has{\mathit{cappuccino}} \\[0.15em]
&& \done{\mathit{pour\_coffee}}\rightarrow\has{\mathit{coffee}} \qquad \done{\mathit{out\_of\_coffee}}\rightarrow\has{\mathit{coffee}} \\[0.15em]
&& \done{\mathit{pour\_tea}}\rightarrow\has{\mathit{tea}} \qquad
\done{\mathit{out\_of\_tea}}\rightarrow\has{\mathit{tea}} \\[0.15em]
&& \done{\mathit{pour\_milk}}\rightarrow\has{\mathit{cappuccino}} \qquad \done{\mathit{out\_of\_milk}}\rightarrow\has{\mathit{cappuccino}} \\[0.25em]
%&& \text{/* Contextual information */}\\ 
\textit{S}_2 & \doteq & 
\mathit{in(Europe)}\\[0.25em]
%&& \text{/* Quantitative constraints */}\\ 
%\textit{S}_3 & \doteq & 
%\mathit{\ldots quantitative\ constraints\ldots}\\[0.5em]
\textit{Q} & \doteq & \textit{D} + (\install{\mathit{pre}\textit{-}\mathit{conf}},10).\textit{R}\\[0.2em]
\textit{D} & \doteq & 
(\install{\mathit{euro}},10).\textit{Q} + (\install{\mathit{dollar}},10).\textit{Q} + (\install{\mathit{sugar}},10).\textit{Q} + (\install{\mathit{cancel}},7).\textit{Q}\\[0.1em]
& & {} + (\install{\mathit{coffee}},9).\textit{Q} + (\install{\mathit{tea}},6).\textit{Q} + (\install{\mathit{cappuccino}},3).\textit{Q}\\[0.2em]
\textit{R} & \doteq & ((\mathit{euro},25).\emptyset + (\mathit{dollar},25).\emptyset); \textit{P}_1\\[0.15em]
\textit{P}_0 & \doteq & (\mathit{return\_coin},10).\textit{R} + (\mathit{no\_return},1).\textit{R}\\[0.15em]
\textit{P}_1 & \doteq & (\mathit{cancel},5).\textit{P}_0 + \textit{P}_2 + \textit{P}_3\\[0.15em]
\textit{P}_2 & \doteq & (\mathit{sugar},15).\emptyset; ((\mathit{pour\_sugar},10).\textit{P}_3 + (\mathit{out\_of\_sugar},2).\textit{P}_1)\\[0.15em]
\textit{P}_3 & \doteq & (\mathit{coffee},20).\textit{P}_4 + (\mathit{tea},12).\textit{P}_5
+ (\mathit{cappuccino},8).\textit{P}_6\\[0.15em]
\textit{P}_4 & \doteq & (\mathit{pour\_coffee},10).\textit{P}_8 + (\mathit{out\_of\_coffee},2).\textit{P}_3\\[0.15em]
\textit{P}_5 & \doteq & (\mathit{pour\_tea},10).\textit{P}_8 + (\mathit{out\_of\_tea},2).\textit{P}_3\\[0.15em]
\textit{P}_6 & \doteq & (\mathit{pour\_milk},10).\emptyset; ((\mathit{pour\_coffee},10).\textit{P}_8 + (\mathit{out\_of\_coffee},2).\textit{R}) + (\mathit{out\_of\_milk},2).\textit{P}_3\\[0.15em]
\textit{P}_8 & \doteq & \textit{P}_9 + (\install{\mathit{ringtone}},8).(\mathit{ringtone},18).\textit{P}_9\\[0.15em]
\textit{P}_9 & \doteq & (\mathit{take\_drink},10).\mathit{R} + (\mathit{no\_cup},1).\textit{R}
\end{array}
$$
\hrulefill
\caption{\pflan{} specification of the family of coffee machines (instantiated for Europe)}
\label{fig:coffee}  
\end{figure}

\enlargethispage*{\baselineskip}
The store, instead, is made up of two parts: constraints derived from the requirements~($\mathit{S}_1$) plus contextual information~($\mathit{S}_2$). % and quantitative constraints ($\mathit{S}_3$). 
The current action constraints are quite simple (all are of the form $\done{\mathit{f}}\rightarrow\has{\mathit{g}}$) but, as said before, they could be more sophisticated upon need (e.g.\ the constraint on action $\mathit{cappuccino}$ could be specified as $\done{\mathit{cappuccino}}\rightarrow\has{\mathit{cappuccino}}\wedge\has{\mathit{ringtone}}$ to require not only the presence of its corresponding feature but also that of the $\mathit{ringtone}$ feature). In Fig.~\ref{fig:coffee}, a product line of European coffee machines is instantiated by the explicit context information $\mathit{in(Europe)}$.

The configuration process~$\mathit{D}$ is a simple rated choice among the installation of some of the features a coffee machine may exhibit. This specifies a sort of race between features and may be thought of as independent designers competing to install the features for which they are responsible. The semantics of \pflan{} ensures that all executions will result in a consistent configuration if the process begins with a consistent store, i.e.\ the semantics forbids the installation of features that are mutually exclusive or prohibited by (a combination of) the constraints. %quantitative constraint.
Formally, multiple installations of the same feature does not have any effect, as installed features are organised in a set.
The rates of the actions influence this race by determining a higher (or lower) probability for the installation of one feature with respect to another (or prior to another).
In our example, to reflect the fact that $\mathit{coin}$ and $\mathit{sugar}$ are core features, we assign higher rates to them than to the optional features to raise their chances of being installed first. Moreover, since we are modelling a coffee machine and since coffee is a necessary ingredient for cappuccino, we assign a higher rate to the feature $\mathit{coffee}$ than to those of other drinks.
As a result, %in our product line 
the probability to install $\mathit{sugar}$ in the first step, given that also $\emph{pre-conf}$, $\mathit{euro}$, %$\mathit{dollar}$, DOLLAR CANNOT BE INSTALLED, AND WE CAN install pre-configured
$\mathit{cancel}$, $\mathit{coffee}$, $\mathit{tea}$ and $\mathit{cappuccino}$ can be installed, thus becomes $\frac{10}{10+10+10+7+9+6+3} = \sfrac{2}{11}$. %, which is in fact much higher than installing $\mathit{tea}$ in the first step: $\frac{3}{10+10+10+7+9+6+3} = \sfrac{3}{55}$.
%We remark here that the actual rates in an SPL model could be  inferred from a statistical analysis of the historical record of SPLs, if available.  

Process~$\mathit{R}$, finally, describes the run-time execution of a coffee machine. The machine may either accept a euro or a dollar, depending on the market it is meant for. After that, the user may cancel coin insertion, upon which the machine returns to its initial state and (usually) returns the coin. With a probability of~$\sfrac{1}{11}$, however, the machine does not return the coin (viz.\ $\frac{1}{10+1}$). If coin insertion is not canceled, the user may ($\textit{P}_2$) or may not ($\textit{P}_3$) push a button for sugar. In case sugar is selected, it is also poured, after which the user can select a beverage. But, with a probability of~$\frac{2}{10+2} = \sfrac{1}{6}$ the machine is out of sugar, after which the user may either cancel the coin insertion or go for an unsugared drink.
Beverage selection (more likely coffee than tea or cappuccino) is followed by the drink being poured (again with a probability that the chosen drink is unavailable), which in case of cappuccino concerns both milk and coffee. 
In case coffee or tea was chosen but unavailable, the user can again choose a beverage (and the machine may have been refilled). 
In the specific case that milk was poured but coffee is not available, the user has bad luck as the machine returns to its idle state before completing the chosen beverage. In case a drink was poured successfully, a ringtone may follow (in which case it first needs to be installed). The user then either takes the drink or, with a~$\sfrac{1}{11}$ probability, realizes that sadly enough there was no cup available. Either way, the machine returns to its initial state. 

Note how the rates \lq influence\rq\ the behavior, in the sense that the choice operator is no longer purely non-deterministic, but probabilistic, i.e.\ the rates provide a probabilistic model of the behavior of the coffee machine and its environment (the users).
Consider, e.g., the choice of a beverage: $(\mathit{coffee},20).\textit{P}_4 + (\mathit{tea},12).\textit{P}_5 + (\mathit{cappuccino},8).\textit{P}_6$. The probability to choose coffee is \sfrac{1}{2} (viz.\ $\frac{20}{20+12+8}$), compared to~$0.3$ for tea and~\sfrac{1}{5} for cappuccino.
Similarly, the probability to cancel coin insertion is~$\frac{5}{5+15+20+12+8} = \sfrac{1}{11}$ (i.e.\ rather low). Note that we need to expand processes~$\textit{P}_2$ and~$\textit{P}_3$ to calculate this probability.
%Intuitively, the rates represent the behavior of the users who are using the coffee machine and of the functioning of the coffee machine itself, allowing for some uncertainty. 

The rates that we assigned in this example merely serve to illustrate the proof-of-concept that we present in this paper. In practice, those rates may be obtained from a statistical analysis of the actual product configuration processes and product behaviours, possibly contained in historical logs. 
% actual user and system behaviour (e.g.\ obtained from data mining a system's log files).
 
%\enlargethispage*{\baselineskip} 
Note that~$\mathit{D}$ and~$\mathit{R}$ are not purely distinct (pre-)configuration and run-time processes, respectively: feature $\mathit{ringtone}$ may be installed dynamically at run time (i.e.\ possibly by~$\mathit{R}$ but never by~$\mathit{D}$) and it can be thought of as, e.g., a software module. This is an example of a staged configuration process, in which some optional features are bound at run time rather than at (pre-)configuration time.

\setlength{\belowcaptionskip}{10pt}
% !TEX root = FMSPLE.tex

\section{Quantitative Analysis with MultiVeStA} 
\label{sec:multivesta}

MultiVeStA~\cite{SV13} is a statistical analysis tool developed and maintained by S.~Sebastio and A.~Vandin. It extends the (distributed) statistical model-checking tools PVeStA~\cite{AM11} and VeStA~\cite{SVA05}, developed at the Department of Computer Science of the University of Illinois at Urbana-Champaign. %, with distributed statistical analysis capabilities. 
Differently from its predecessors, MultiVeStA can easily be integrated with any formalism which allows for probabilistic simulations. It has so far been used to analyse transportation systems~\cite{DBLP:conf/ifm/GilmoreTV14}, volunteer clouds~\cite{SEAMS2014}, crowd-steering~\cite{MultivestaAlchemist} and swarm robotic~\cite{MisscelAndPirlo} scenarios. % modelled according to various system specification languages.

In this paper, we use MultiVeStA to analyse \pflan{} specifications %by performing probabilistic simulations 
in order to obtain statistical estimations of quantitative properties expressed in MultiVeStA's query language MultiQuaTEx (an extension of QuaTEx~\cite{AMS06}). MultiVeStA provides such estimations by means of %efficient 
distributed statistical analysis techniques known from statistical model checking~\cite{LDB10,LL14}. 
A prototypical tool integrating MultiVeStA and \pflan{} is available at %\url{www.dropbox.com/s/2zi0m6b5rfx9nfd/distrFMSPLE2015.zip}
\url{https://code.google.com/p/multivesta/wiki/PFLan} 
together with all files necessary to reproduce the experiments discussed in this section.

Probabilistic simulations of a \pflan{} specification can easily be obtained by executing the model step-by-step by applying the rules of Fig.~\ref{fig:semantics}, each time selecting one of the computed one-step next-states according to the probability distribution obtained after normalising the rates of the generated transitions. 
Classical statistical model checking techniques allow one to perform analyses like \lq\lq is the probability that a property holds greater than~$0.3$?\rq\rq\ or \lq\lq what is the probability that a property is satisfied?\rq\rq\ over a given specification. Next to performing such kinds of analyses over products, MultiVeStA also allows to estimate the expected values of properties that can take on any value from~$\real$, like \lq\lq what is the average cost of products  %product configurations
generated from a software product line specification?\rq\rq. %It does so by evaluating the property expressed in MultiQuaTEx with respect to~$n$ independent simulations, with~$n$ large enough to respect a user-specified confidence interval. Confidence intervals are specified in terms of two parameters:  $\alpha$ and $\delta$. 
Estimations are computed as the mean value of $n$ samples obtained from $n$ simulations, with $n$ large enough to grant that the size of the $(1 - \alpha)\times 100\%$ \textit{confidence interval} (CI) %for the expected value 
is bounded by $\delta$. In other words, if a MultiQuaTEx expression is estimated as $\overline{x}$, then with probability $(1 - \alpha)$ its actual expected value belongs to the interval $[\overline{x} - \sfrac{\delta}{2},\linebreak \overline{x} + \sfrac{\delta}{2}]$.  
A CI is thus specified in terms of two parameters: $\alpha$ and $\delta$.
In all experiments discussed in this section, we fixed $\alpha=0.1$, and $\delta=0.1$ and~$\delta=0.5$ for probabilities and costs of products, respectively.
%

% In particular, the user specifies two parameters: $\alpha$ and $\delta$, and MultiVeStA evaluates properties such that if it returns $\overline{x}$, then with probability $1-\alpha$ the actual value of the property is contained in $[\overline -\sfrac{\delta}{2},\overline +\sfrac{\delta}{2},]$

MultiVeStA's property specification language MultiQuaTEx is very flexible, based on the following ingredients: real-valued observations on the current \lq state\rq\ (e.g.\ the total cost of installed features), arithmetic expressions and comparison operators, if-then-else statements, a one-step next operator (which triggers the execution of one step of a simulation) and recursion.
Intuitively, we can use MultiQuaTEx to associate a value from~$\real$ to each simulation and subsequently use MultiVeStA to estimate the expected value of such number (in case this number is~$0$ or~$1$ upon the occurrence of a certain event, we thus estimate the probability of such an event to happen).

\enlargethispage*{\baselineskip}
%\paragraph*{Example}
\section{Quantitative Analyses of the Example Product Line}
\label{sec:analysis}
%\paragraph*{Quantitative Analysis of the Example Product Line.}
%
%\medskip 
Some properties that we can verify over our toy example are as follows: 
\begin{enumerate}%\itemsep=-1pt\parsep=-1pt
\item[$P_1$]
%($P_1$) 
The probability to run into a deadlock before completing the pre-configuration phase; 
\item[$P_2$]
%($P_2$) 
For each of the $8$ primitive features (sugar, ringtone, cancel, cappuccino, coffee, tea, dollar, euro), the probability to have it installed after the pre-configuration phase or at a given simulation step $x$; 
\item[$P_3$]
%($P_3$)
The average cost of products obtained from the pre-configuration phase, or of the \lq intermediate\rq\ ones obtained at a given simulation step $x$.
\end{enumerate}
Note that we consider any configuration obtained by intermediate steps to be a (possibly intermediate) product. This may thus refer to an unfinished product or to underspecified software, or concern a not yet fully developed product. When no more features can be installed, we speak of a final product.
%\marginpar{make use of $\install{\mathit{pre}\textit{-}\mathit{configured}}$}

While not explicitly stated, all experiments discussed in this section refer to versions (defined below) of the \pflan{} specification of Fig.~\ref{fig:coffee} without the contextual information $\textit{S}_2  \doteq \mathit{in(Europe)}$, so as to study properties of our example without restrictions to a specific context (thus implicitly allowing deadlocks).

Property~$P_1$ is useful for studying the correctness of the \pflan{} specification of a product line, in this case by verifying the probability to successfully complete the pre-configuration phase of a product from the product line. 
Property~$P_2$ is useful for studying how often (on average) a feature is actually installed in a product from the product line, which is important information for those designers or programmers responsible for the production or programming of a specific feature or software module.
Property~$P_3$, finally, is useful for studying the average cost of assembling a product from the product line, based on the costs of the features constituting a product defined by the attributed feature model depicted in Fig.~\ref{fig:AFM}.

%\medskip

\begin{figure}[b!]
%\vspace{0.4cm}
\begin{center}
%\vspace*{0.25cm}
\hrulefill 
\vspace*{-0.25cm}
$$
\begin{array}{rcl}
F' & \doteq & [S\mid D'; (\install{\mathit{pre}\textit{-}\mathit{conf}},10).R]\\[0.2em]
D' & \doteq & 
(\install{\mathit{euro}},10).\emptyset \parallel (\install{\mathit{dollar}},10).\emptyset \parallel (\install{\mathit{sugar}},10).\emptyset \parallel (\install{\mathit{cancel}},7).\emptyset\\[0.1em]
& & {} \parallel (\install{\mathit{coffee}},9).\emptyset \parallel (\install{\mathit{tea}},6).\emptyset \parallel (\install{\mathit{cappuccino}},3).\emptyset\\
\end{array}
$$

\vspace*{-0.25cm}
\hrulefill
\caption{\label{fig:coffeeVariant}A modified version of the \pflan{} specification of Fig.~\ref{fig:coffee}}
\end{center}
\end{figure}

Listing~\ref{listing:Q1} depicts a MultiQuaTEx expression to evaluate~$P_1$. \llines{line:tempName}{line:endif1} define a recursive temporal operator which is evaluated against a simulation: it gives~$0.0$ if the feature $\emph{pre-conf}$ is installed in the current simulation state (\lline{line:beginif1}); it gives~$1.0$ if the current state is a deadlock (\lline{line:beginif2}); or it is recursively evaluated in the next simulation state (\lline{line:next}). Intuitively, $\#$ is the one-step temporal operator, while real-valued observations on the current state are evaluate resorting to the keyword \texttt{s.rval}. A number of predefined observations is currently supported, e.g.\ we can query whether a given feature is currently installed (as in \lline{line:beginif1} for $\emph{pre-conf}$) or whether the current process has no more actions that are allowed by the constraints, in which case we say that it is in a deadlock state (\lline{line:beginif2}).  
Finally, \lline{line:e} specifies the property to be studied: the expected value of the defined recursive temporal operator.\\

\lstset{caption=The MultiQuaTEx expression corresponding to property $P_1$, label=listing:Q1}
\begin{lstlisting}[mathescape,float=h,numbers=left]  
DeadlockInPreconf() =@\label{line:tempName}@
 if {s.rval("pre-conf") == 1.0} then 0.0  @\label{line:beginif1}@
 else if {s.rval("deadlock") == 1.0} then 1.0 @\label{line:beginif2}@
 else $\textbf{\#}$DeadlockInPreconf()@\label{line:next}@ fi fi ;@\label{line:endif1}@
 eval E[ DeadlockBeforePreconf() ] ;@\label{line:e}@
\end{lstlisting}
%	                     	     fi
% fi ;@\label{line:endif1}@
% DeadlockInPreconf() = if {s.rval("pre-conf") == 1.0} 
%                          then 0.0  
%                          else if {s.rval("deadlock") == 1.0} 
%                                  then 1.0 
%                                  else $\textbf{{\#}}$DeadlockInPreconf() 
%	                       fi
%                       fi ;

%\enlargethispage*{\baselineskip}
\noindent
We evaluated $P_1$ against our \pflan{} model, %the model of Fig.~\ref{fig:coffee}, 
obtaining probability $0.0$, i.e. the pre-configuration phase (almost surely) always terminates. 

\enlargethispage*{\baselineskip}
Now consider our model to be % of Fig.~\ref{fig:coffee} 
modified according to Fig.~\ref{fig:coffeeVariant}, i.e.\ by replacing $F$ with $F'$, and both $Q$ and $D$ with $D'$. This version still contains a pre-configuration phase ($D'$) followed by the same run-time phase ($R$) of the original model. Essentially, $D'$ tries to install all features, possibly in different orders.\\

By evaluating~$P_1$ against the modified version of our model we obtain probability~$1.0$, i.e.\ the pre-configuration phase (almost surely) never terminates. In fact, we can install only one among $\mathit{dollar}$ or $\mathit{euro}$ (cf.\ the first constraint of $S_1$ in Fig.~\ref{fig:coffee}), and consequently one of the two installations will never succeed. 
$P_1$ can thus indeed be used to check liveness properties of \pflan{} specifications, e.g.\ to individuate %inconsistent 
specifications leading, with a certain probability, to deadlocks. 

Listing~\ref{listing:Q23} depicts a MultiQuaTEx expression to evaluate~$P_2$ and~$P_3$ when considering the products obtained after the pre-configuration phase. 
Such an expression shows how MultiQuaTEx allows one to express more properties at once, which can be estimated by MultiVeStA reusing the same simulations. 
\llines{line:tempName1}{line:tempName1end} define the recursive temporal operator \texttt{ProductCostAfterPreconf}. It is evaluated against a simulation as the cost of the product obtained from the pre-configuration phase. As shown in \lline{line:costObs}, a further predefined observation is supported, viz.\ \emph{cost\/}, which provides the cost of the current product. 
\llines{line:tempName2}{line:tempName2end} define a parametric recursive temporal operator which evaluates to~$1.0$ if the feature provided as parameter is installed during the pre-configuration phase, and to~$0.0$ otherwise.
Finally, \llines{line:evalBegin}{line:evalEnd} specify the properties to be analysed: the average cost of products generated by the pre-configuration phase (\lline{line:evalBegin}) and for each of the~$8$ primitive features the probability to have it installed (\llines{line:evalBegin2}{line:evalEnd}).
We remark that MultiVeStA adopts a procedure which takes into account that each property might require a different number of simulations to satisfy the required confidence interval CI.\\

\lstset{caption=The MultiQuaTEx expression corresponding to properties $P_2$ and $P_3$, label=listing:Q23}
\begin{lstlisting}[mathescape,float=h,numbers=left]     
ProductCostAfterPreconf() = @\label{line:tempName1}@
 if {s.rval("pre-conf") == 1.0} then s.rval("cost")  @\label{line:costObs}@
                                else $\textbf{\#}$ProductCostAfterPreconf() fi ;@\label{line:tempName1end}@
IsInstalledAfterPreconf(feature) =  @\label{line:tempName2}@
 if {s.rval("pre-conf") == 1.0} then s.rval(feature)
                                else $\textbf{\#}$IsInstalledAfterPreconf({feature}) fi ;@\label{line:tempName2end}@
eval E[ ProductCostAfterPreconf() ]; @\label{line:evalBegin}@
eval E[ IsInstalledAfterPreconf("sugar") ]; eval E[ IsInstalledAfterPreconf("ringtone") ]; @\label{line:evalBegin2}@
eval E[ IsInstalledAfterPreconf("cancel") ]; eval E[ IsInstalledAfterPreconf("cappuccino") ]; 
eval E[ IsInstalledAfterPreconf("coffee") ]; eval E[ IsInstalledAfterPreconf("tea") ]; 
eval E[ IsInstalledAfterPreconf("dollar") ]; eval E[ IsInstalledAfterPreconf("euro") ];@\label{line:evalEnd}@
\end{lstlisting}

\enlargethispage*{\baselineskip}
\noindent
We evaluated the MultiQuaTEx expression of Listing~\ref{listing:Q23} against the original model of Fig.~\ref{fig:coffee}. The obtained average cost is $14.07$, %$14.47$, 
while the probabilities of installing the primitive features are given in the first row of Table~\ref{table:Q2}.
Clearly, the probability with which features are installed (as well as the average cost of the obtained products) is highly affected by the rate at which $\emph{pre-conf}$ is installed ($10$ in Fig.~\ref{fig:coffee}): a lower or higher rate rate leads to more or less iterations of $\mathit{D}$, respectively. 
In order to quantify the influence of this rate, we further evaluated the expression of Listing~\ref{listing:Q23} against the model obtained changing the aforementioned rate to %$20$ and 
$50$. The obtained average cost of products is $4.46$, while the probabilities of installing the features are provided in the second row of Table~\ref{table:Q2}. As expected, the higher installation rate of $\emph{pre-conf}$ has the effect of decreasing the average number of iterations of the pre-configuration phase, leading to a lower probability of installation of the features and to a lower average cost of products.

\begin{table}[h!]
\vspace*{0.3cm}
\centering 
\scalebox{0.99}{
\begin{tabular}{rcccccccc}
\toprule
%\emph{Feature~~} &
   &  \multicolumn{8}{c}{Features}
 \\
    \cmidrule(l){2-9}
\multicolumn{1}{r}{Rate of $install(\emph{pre-conf})$~~} & \multicolumn{1}{c}{$\mathit{sugar}$} &  \multicolumn{1}{c}{$\mathit{ringtone}$} & \multicolumn{1}{c}{$\mathit{cancel}$} &
\multicolumn{1}{c}{$\mathit{cappuccino}$} & \multicolumn{1}{c}{$\mathit{coffee}$} & \multicolumn{1}{c}{$\mathit{tea}$} & \multicolumn{1}{c}{$\mathit{dollar}$} &
\multicolumn{1}{c}{$\mathit{euro}$} \\
\midrule
%\emph{Result~~} & 
10~~~~~~~~& 0.49 & 0.0 & 0.45 & 0.13 & 0.50 & 0.40 & 0.33 & 0.38 \\
%\midrule
%20~~& \color{red} XXX & \color{red} XXX & \color{red} XXX & \color{red} XXX & \color{red} XXX & \color{red} XXX & \color{red} XXX & \color{red} XXX \\
%\midrule
50~~~~~~~~&  0.17 & 0.0 & 0.11 & 0.0 & 0.14 & 0.10 & 0.12 & 0.13 \\
\bottomrule
\end{tabular} 
}
\caption{\label{table:Q2}Probability of installing a feature during the pre-configuration phase of model in Fig.~\ref{fig:coffee} ($P_2$).}
%\vspace*{\baselineskip}
\end{table}

We conclude this section by showing how MultiVeStA can be used to analyse properties of a \pflan{} specification upon varying the number of performed simulation steps.
Listing~\ref{listing:Q23Param} sketches how the MultiQuaTEx expression of Listing~\ref{listing:Q23} can be made parametric with respect to a given set of simulation steps. 
First of all, the temporal operators were modified so that they are evaluated with respect to a specific step given as parameter. We actually provide only the updated temporal operator regarding the costs of products (\llines{line:tempName1Param}{line:tempName1endParam}), as the other has been modified similarly.
Subsequently, it is necessary to specify a range of values for the parameter. \llines{line:evalParam1}{line:evalParam2} specify that we are interested in studying the properties for steps going from~$0.0$ to~$40$, with an increment of~$1.0$.\\

\lstset{morekeywords=parametric,caption=The MultiQuaTEx expression corresponding to $P_2$ and $P_3$ upon varying the simulation steps, label=listing:Q23Param}
\begin{lstlisting}[mathescape,float=h,numbers=left]     
ProductCostAtStep(step) = @\label{line:tempName1Param}@
 if {s.rval("steps") == step} then s.rval("cost")  @\label{line:costObsParam}@
                              else $\textbf{\#}$ProductCostAfterPreconf(step)
 fi ;@\label{line:tempName1endParam}@
IsInstalledAtStep(step,feature) =  $\ldots$
eval parametric(E[ ProductCostAtStep(step) ],E[ IsInstalledAtStep(step,"sugar") ],$\ldots$, @\label{line:evalParam1}@
                E[ IsInstalledAtStep(step,"euro") ],step,0.0,1.0,40.0) ;@\label{line:evalParam2}@
\end{lstlisting}

\begin{figure}[h!]
\begin{center}
\includegraphics[width=0.9\textwidth]{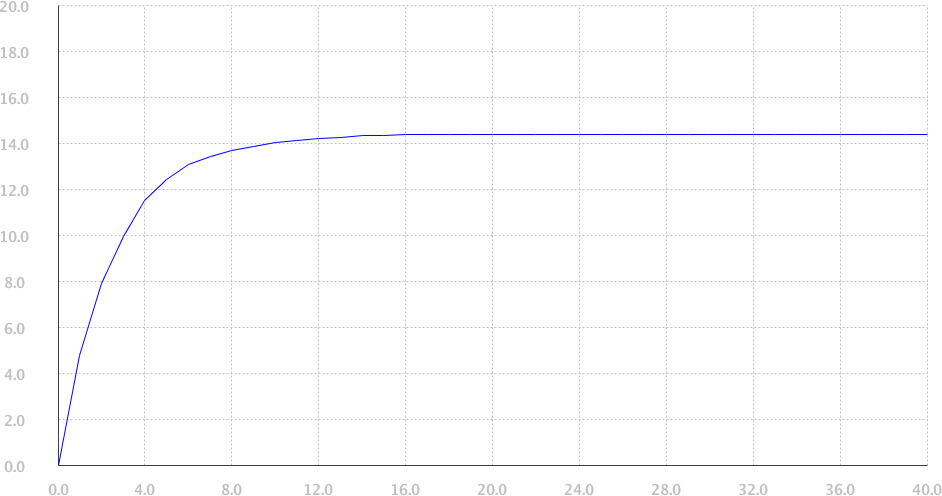}
\caption{Average product cost during the first $40$ steps}% analysis result of $P_3$ 
\label{fig:parametricCosts}
\end{center}
\end{figure}

\begin{figure}[hb!]
\begin{center}
\includegraphics[width=0.9\textwidth]{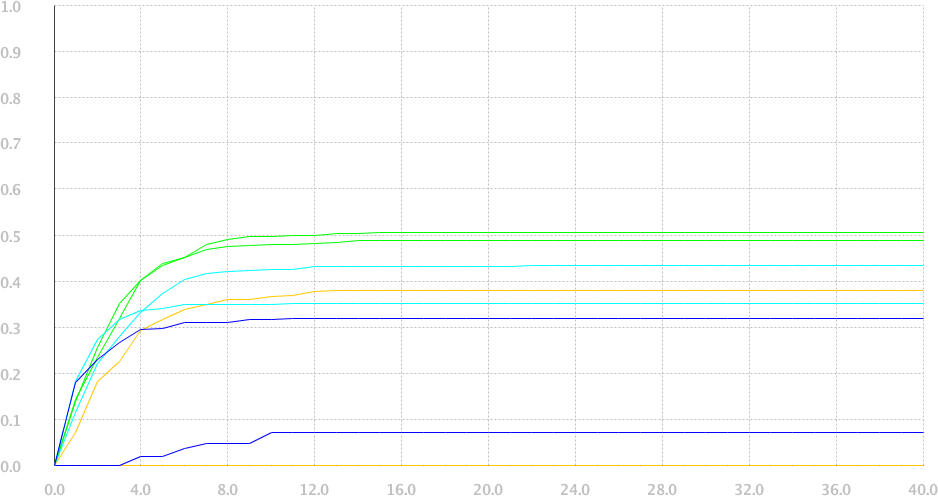}
\caption{Average feature installation probability during the first $40$ steps}% analysis result of $P_3$ 
\label{fig:parametricInstallationProbs}
\end{center}
\end{figure}

\noindent
We evaluated also the parametric property of Listing~\ref{listing:Q23Param} against the original model of Fig.~\ref{fig:coffee}. All such analyses ($41\times 9$ different properties) were evaluated using the same simulations. The results are presented in two plots: one for costs (\texttt{ProductCostAfterPreconf}) %on the left-hand side of Fig.~\ref{fig:parametricCostsAndInstallationProbs} 
in Fig.~\ref{fig:parametricCosts}, and one for probabilities (\texttt{IsInstalledAfterPreconf}) %on the right-hand side of Fig.~\ref{fig:parametricCostsAndInstallationProbs}. 
in Fig.~\ref{fig:parametricInstallationProbs}.\\ 

\enlargethispage*{\baselineskip}
\vspace*{-0.25cm}\noindent
As expected, %Fig.~\ref{fig:parametricCostsAndInstallationProbs} (left)
Fig.~\ref{fig:parametricCosts} shows that the average cost (on the y-axis) of the intermediate products generated from the software product line grows with respect to the number of performed simulation steps. 
In particular, it shows a fast growth during the first $6$ steps, reaching an average cost of~$13$, and then it essentially stabilises, eventually reaching its maximum~($14.38$) from step~$26$ onwards. This is consistent with our \pflan{} specification, %given in Fig.~\ref{fig:coffee},
consisting of a pre-configuration phase during which the majority of the features are installed, followed by a run-time phase modelling the behaviour of the generated product (and possibly installing $\mathit{ringtone}$).

%Figure~\ref{fig:parametricCostsAndInstallationProbs} (right) 
Fig.~\ref{fig:parametricInstallationProbs} shows that the probabilities (on the y-axis) for each of the features to be installed evolve similarly to the average cost of the generated products, although, clearly, with different scales: they show a fast growth during the first $6$ steps, after which they essentially stabilise while approaching their maximum. The maximum probabilities of installing the various primitive features are as follows: $0.0$~for $\mathit{ringtone}$, $0.07$~for $\mathit{cappuccino}$, $0.32$~for $\mathit{euro}$, $0.35$~for $\mathit{dollar}$, $0.38$~for $\mathit{tea}$, $0.43$~for $\mathit{cancel}$, $0.49$~for $\mathit{coffee}$ and $0.51$~for $\mathit{sugar}$. 
%
%It is worth to note that t
Note that the probability of installing $\mathit{ringtone}$ is really very low ($0.0$ or, to be precise, its actual expected value belongs to the interval $[0, 0 + \sfrac{\delta}{2}] = [0, 0 + \frac{0.1}{2}] = [0, 0.05]$). Indeed, while the installation of $\mathit{ringtone}$ is allowed by our specification, it is optional (except when serving cappuccino). 
%This is reflected in the very low computed probability. 
Note, however, that we considered simulations consisting of only $40$ steps (the x-axes in the two figures). For longer simulations, we would of course have obtained a higher probability to install $\mathit{ringtone}$.\linebreak

\vspace*{0.25cm}
%Fig.~\ref{fig:parametricCostsAndInstallationProbs} (right) shows that the probabilities of having with which features are installed respect the different rates provided in Fig.~\ref{fig:coffee} for their installation \ldots

%\begin{figure*}
%\begin{center}
%\includegraphics[width=0.475\textwidth]{analysis/costAndProbabilitiesOfInstallingParametric/parametricCostsCropped.png}
%\hspace{0.25cm}
%\includegraphics[width=0.475\textwidth]{analysis/costAndProbabilitiesOfInstallingParametric/parametricFeaturesInstallingProbabilitiesCropped.png}
%\caption{Average product cost (left) and feature installation probability (right) during the first $40$ steps}% analysis result of $P_3$ 
%\label{fig:parametricCostsAndInstallationProbs}
%\end{center}
%\end{figure*}

%\begin{figure}
%\begin{center}
%\includegraphics[width=0.7\textwidth]{analysis/costAndProbabilitiesOfInstallingParametric/parametricCosts.png}
%\caption{The average costs of productsin each of the first $40$ steps of simulation.}
%\label{fig:parametricCosts}
%\end{center}
%\end{figure}
%
%\begin{figure}
%\begin{center}
%\includegraphics[width=0.7\textwidth]{analysis/costAndProbabilitiesOfInstallingParametric/parametricFeaturesInstallingProbabilities.png}
%\caption{The probability of installing each feature in each of the first $40$ steps of simulation.}
%\label{fig:parametricInstallationProbs}
%\end{center}
%\end{figure}
%\input{RW.tex}
% !TEX root = FMSPLE.tex

\section{Conclusion}
\label{sec:discussion}

In this paper, we have continued a line of research presented at earlier editions of FMSPLE~\cite{GP12,BLP13} by enriching \flan{}, a high-level feature-based modelling language for software product lines, with quantitative information. The result, \pflan{}, allows one to model and analyse the likelihood of installing features, the probabilistic behaviour of users of products of the product line, and the costs of products, next to probabilistic quantifications of ordinary temporal properties (e.g.\ \lq\lq what is the probability that coffee is poured while no cup was available?\rq\rq).
In addition, we extended the qualitative analysis framework for software product lines implemented in Maude with statistical techniques for quantitative analysis.

The modelling and analysis capabilities of \pflan{} were illustrated on a simple product line of coffee machines. 
In the future, we plan to investigate the scalability of our (tool) framework by considering more realistic and complex scenarios.
We also intend to add the possibility to define quantitative constraints to \pflan{}, possibly by adopting further operations from extensions of the concurrent constraint paradigm that can deal with quality of service and mobility~\cite{BM07} and its stochastic extension~\cite{Bor06}. 
Both the \textsf{check} operation of concurrent constraint programming, to prevent inconsistencies, and its \textsf{retract} operation, to remove (syntactically present) constraints from the store, might be useful to enable the dynamic (un)installation of features in the presence of (\emph{soft}) quantitative constraints (i.e.\ not only Boolean~\cite{CSHL13}).
In particular, we would like to investigate the consequences of allowing the explicit uninstallation of a feature, e.g.\ due to its malfunctioning or due to the need of replacing it by a better (version of the) feature.
Such features were shown successful in services computing for the specification of service-level agreements and negotiation processes~\cite{BM11}. %Moreover, in dynamic software product lines it is common to consider staged configuration processes in which some features are bound at run time rather than at (pre-)configuration time~\cite{CHE04,BLLBGS14}. 
Finally, we would like to allow behaviour that is explicitly influenced by the constraint store, as in~\cite{Bor06}. In our example, this would allow us to model, e.g., the probability of a user choosing a coffee to depend on the location of the coffee machine (i.e.\ Europe or Canada), thus allowing us to assign, e.g., a higher weight to ordering cappuccino in Europe.

\subsection*{Acknowledgements}

This research was supported by the EU FP7-ICT FET-Proactive project QUANTICOL (600708) %, by ... other projects? ..., 
and the Italian MIUR project CINA (PRIN 2010LHT4KM).

Moreover, we thank the reviewers for their detailed comments, which helped us to improve the paper.

%%%%%%%%%%%%%%%%%%%%%%%%%%%%%%%%%%%%%%%%%
%%%%%%%%%%%%%%%% NOTES %%%%%%%%%%%%%%%%%%%%%
%%%%%%%%%%%%%%%%%%%%%%%%%%%%%%%%%%%%%%%%% 
%\smallskip
%
%Some further ideas (postpone to the paper for SPLC?):
%%%
%\begin{enumerate}
%\item Once we have implemented the possibility to specify quantitative constraints over features with non-Boolean attributes in \pflan{}, we %can exploit our implementation of Microsoft's Z3 SMT constraint solver~\cite{MB08} in Maude to optimize the process of feature %configuration.
%%%Selecting or not an expensive feature heavily influences the choice of other features. 
%\item Can we set the cost of a feature to a variable, say $x$, and then run such an optimization to find the best value for $x$?
%\item Can we actually make the costs of features visible in a product family's behaviour and consequently control/optimize (product) %behaviour at run time?
%\end{enumerate}
%
%%%how does the dynamic installation work when there's a cost function to consider?!

%%%%%%%%%%%%%%%%%%%%%%%%%%%%%%%%%%%%%%%%%
%%%%%%%%%%%%%%%% NOTES %%%%%%%%%%%%%%%%%%%%%
%%%%%%%%%%%%%%%%%%%%%%%%%%%%%%%%%%%%%%%%%

\bibliographystyle{eptcs}
\bibliography{FMSPLE}

\end{document}